\documentclass[preprintnumbers,prl,twocolumn,floatfix,preprintnumbers,amsmath,amssymb,nofootinbib,superscriptaddress]{revtex4}
\usepackage{graphicx}
\usepackage{enumitem}
\usepackage{epsfig}
\usepackage{bm}
\usepackage{amsfonts}
\usepackage{color}
\usepackage{xspace}
\usepackage{subfigure}
\usepackage{amsmath}
\usepackage{amssymb}
\usepackage{graphicx, comment}
\usepackage{makecell}
\usepackage{float}
\usepackage[normalem]{ulem}
\usepackage{xcolor}

\usepackage[utf8]{inputenc}
\usepackage[english]{babel}

\usepackage[
      colorlinks=true,
      linkcolor=blue,
      urlcolor=magenta,
      filecolor=green,
      citecolor=red,
      pdfstartview=FitV,
      pdftitle={},
        pdfauthor={Ruchika},
        pdfsubject={},
        pdfkeywords={},
        pdfpagemode=None,
        bookmarksopen=true
      ]{hyperref}
\usepackage{color}

\newcommand{\bcm}{}

\newcommand{\Hunit}{km s$^{-1}$ Mpc$^{-1}$}

\usepackage{amsmath,amssymb}
\date{Accepted XXX. Received YYY; in original form ZZZ}

\begin{document}

\title{Hubble Tension and the G-step Model:\\ Re-examination of Recent Constraints on Modified Local Physics}

\author{Leandros Perivolaropoulos}
\email{leandros@uoi.gr}
\affiliation{Department of Physics, University of Ioannina, GR-45110, Ioannina, Greece}

\author{Ruchika}
\email{ruchika.science@usal.es}
\affiliation{Departamento de Física Fundamental and IUFFyM, Universidad de Salamanca, E-37008 Salamanca, Spain}

\date{\today}

\begin{abstract}
We critically examine recent claims challenging the viability of the G-step model (GSM) as a solution to the Hubble tension. The GSM proposes a $\sim$4\% increase in the effective gravitational constant $G_{\text{eff}}$ beyond $z \approx 0.01$ to reconcile local and early-universe measurements of the Hubble constant. Through detailed quantitative analysis, we demonstrate that many proposed constraints on the model require careful reconsideration. Key findings include: (1) Modern stellar modeling indicates a weaker $L \propto G^4$ scaling rather than the traditional $G^7$, significantly reducing tension with stellar evolution constraints; (2) The fluid-like behavior of Earth 150 Myr ago preserves the day/year ratio across any $G$ transition; (3) Paleoclimate data showing $\sim$20°C cooling over relevant timescales appears consistent with, rather than challenging, the GSM; (4) Distance indicator comparisons allow for $\Delta G/G$ variations up to $\sim$20\% at 2$\sigma$ when systematic uncertainties are properly included; (5) The discrete nature of the proposed $G$ transition preserves relative stellar population ages used in cosmic chronometry. When accounting for proper uncertainty levels in both observations and theoretical modeling, we find the GSM remains a viable candidate for resolving the Hubble tension. We identify specific observational tests with next-generation facilities that could definitively confirm or rule out the model.
\end{abstract}



\maketitle
\section{Introduction}
\subsection{The Hubble Tension: A Brief Overview}

The Hubble constant ($H_0$) is a fundamental cosmological parameter that quantifies the current expansion rate of the Universe. Its precise determination is crucial for understanding the Universe's age, composition, and evolution. Despite significant advances in observational techniques and theoretical modeling, a persistent discrepancy known as the "Hubble tension" has emerged between different measurement methods \cite{Perivolaropoulos:2024yxv,Abdalla:2022yfr}.

This tension is traditionally viewed as a disagreement between early and late-time measurements of $H_0$. Early-time measurements, which use the sound horizon at recombination as a standard ruler, consistently yield $H_0 = 67.4 \pm 0.5$ km/s/Mpc \cite{Planck:2018vyg}. In contrast, late-time measurements based on the cosmic distance ladder, particularly the SH0ES collaboration's analysis, find $H_0 = 73.04 \pm 1.04$ km/s/Mpc \cite{Riess:2021jrx}, suggesting a significantly faster expansion rate.

However, recent analyses \cite{Perivolaropoulos:2024yxv} indicate that this characterization may be oversimplified. The tension appears to be more accurately described as a discrepancy between distance ladder measurements and all other methods, including both early-time measurements and local one-step measurements that bypass the distance ladder approach. Distance ladder measurements consistently yield higher values ($H_0 \approx 72.8 \pm 0.5$ km/s/Mpc), while one-step measurements, independent of both the CMB sound horizon and the distance ladder, suggest lower values ($H_0 \approx 69.0 \pm 0.5$ km/s/Mpc).

This reframing of the Hubble tension suggests two possible explanations:
\begin{enumerate}
\item A fundamental change in physics affecting the relationships between different rungs of the cosmic distance ladder \cite{Marra:2021fvf,Perivolaropoulos:2021bds}
\item Unidentified systematic effects that similarly influence all distance ladder calibrators \cite{Efstathiou:2020wxn,Mortsell:2021nzg}
\end{enumerate}

The distance ladder approach relies crucially on the assumption that both the astrophysical environment and physical laws remain consistent across its three rungs. Any violation of this assumption could systematically affect all distance ladder measurements while leaving one-step measurements unaffected. This perspective suggests that resolving the Hubble tension may require closer examination of local physics and potential systematic effects in distance ladder calibration, rather than modifications to early-universe physics as commonly proposed \cite{Poulin:2018cxd,Kamionkowski:2022pkx}.

Various solutions involving modified early-universe physics have been proposed, typically aiming to reduce the sound horizon scale and thereby increase the $H_0$ value derived from CMB observations \cite{Poulin:2018cxd,Smith:2020rxx}. However, these approaches face significant challenges \cite{Jedamzik:2020zmd,Vagnozzi:2021gjh} and may not address the fundamental issue if the tension primarily exists between distance ladder and non-distance ladder measurements rather than between early and late-time observations.

The sound horizon scale, \( r_s \), at recombination (the standard ruler) and the standardized bolometric absolute magnitude \( M_B \) of Type Ia supernovae (SnIa standard candles) are essential tools for measuring cosmological distances and, consequently, the Hubble constant \( H_0 \), one of the most fundamental parameters in cosmology. However, a significant discrepancy exists between the \( H_0 \) values obtained using these two methods, reaching a \( 5\sigma \) tension. This issue challenges the \(\Lambda\)CDM model, the current standard cosmological framework.

To address the Hubble tension, three principal classes of models have been proposed:

1. \textbf{Early-time models:} These models propose new physics beyond the standard \(\Lambda\)CDM paradigm to address the Hubble tension by modifying the conditions and dynamics of the early universe, specifically during the epoch of recombination. Among the most prominent approaches are Early Dark Energy (EDE) \cite{Poulin:2018cxd, Kamionkowski:2022pkx, Simon:2022adh, Braglia:2020bym, Niedermann:2020dwg} and modified gravity theories \cite{Braglia:2021, Brax:2014, Adi:2021, Clifton:2012, Lin:2019, DiValentino:2016}. These approaches are designed to alter the sound horizon scale \( r_s \), which is the comoving distance that acoustic waves in the photon-baryon plasma travel prior to decoupling. The sound horizon plays a critical role in calibrating the standard ruler used in cosmology, and its size directly impacts the inference of the Hubble constant \( H_0 \). By decreasing \( r_s \), these models aim to reconcile the \( H_0 \) value inferred from early-universe observations, such as the Cosmic Microwave Background (CMB), with that derived from late-universe measurements, such as Cepheid-calibrated Type Ia supernovae.

Despite the promise of these approaches, early-time models face several challenges. One of the primary issues is that while they can reduce the Hubble tension, they often fail to fully resolve it. For example, EDE models typically predict an \( H_0 \) value that, although higher than the standard \(\Lambda\)CDM prediction, still falls short of the value measured from local observations. Furthermore, these models often exacerbate the \( S_8 \) tension, which pertains to the discrepancy between the amplitude of matter clustering (\( \sigma_8 \)) inferred from CMB data and that measured through weak lensing and large-scale structure surveys. This arises because changes in the early-universe expansion rate can lead to enhanced growth of matter perturbations, increasing the clustering amplitude and creating tension with large-scale structure data.

Another major challenge is the need for fine-tuning. Both EDE and modified gravity models require precise adjustments to their parameters to achieve compatibility with observational data. For EDE, this includes the initial conditions and potential of the scalar field, as well as the epoch and duration of its energy contribution. For modified gravity, the additional degrees of freedom introduced must be carefully constrained to avoid observable deviations in other cosmological and astrophysical phenomena, such as gravitational lensing, galaxy clustering, or the propagation of gravitational waves.

2. \textbf{Late-time models (H(z) deformation):} These models propose deformations in the Hubble expansion history \( H(z) \) at low redshifts (\( z \lesssim 2 \)) as a potential solution to reconcile the discrepancy in \( H_0 \) values derived from early-universe observations, such as the Cosmic Microwave Background (CMB), and late-universe measurements, such as Type Ia supernovae (SnIa) \cite{Li:2019, Pan:2020, Panpanich:2021, DiValentino:2020a, DiValentino:2020b, Li:2020, Clark:2021, DiValentino:2020c}. The idea is to modify the behavior of the universe's expansion at relatively recent times to adjust the inferred Hubble constant without altering early-universe physics. 

However, these models face significant challenges. One major issue is their difficulty in simultaneously fitting the Baryon Acoustic Oscillation (BAO) data, which is sensitive to the sound horizon scale \( r_d \), and the SnIa data, which directly probes the expansion history \cite{Bousis:2024rnb}. This tension arises because modifying \( H(z) \) at late times can introduce inconsistencies in the distance ladder calibration or lead to deviations in the angular diameter distance measurements at different redshifts.

Additionally, recent results from the Dark Energy Spectroscopic Instrument (DESI) survey provide new constraints on the equation of state parameter \( w \), which characterizes the behavior of dark energy. DESI data suggest that the equation of state \( w \) must remain in a non-phantom regime (\( w \geq -1 \)), indicating that dark energy does not behave as a "phantom" fluid with \( w < -1 \) \cite{DESI:2024mwx}. This finding is in contrast to earlier analyses, where other datasets either favored a cosmological constant (\( w = -1 \)) consistent with the $\Lambda$CDM model or allowed \( w \) to dip below \(-1\), suggesting a phantom state.

The requirement for \( w \) to remain non-phantom complicates late-time models further, as many proposed modifications to the expansion history rely on phantom-like dark energy to accelerate the universe's expansion more strongly at late times. Balancing the DESI constraints with other datasets, while maintaining consistency with observed BAO and SnIa data, presents a significant theoretical challenge. This highlights the difficulty in constructing late-time models that are both observationally viable and capable of resolving the Hubble tension in a natural and consistent manner.

3. \textbf{Ultralate-time models:} These models propose environmental or physical changes, such as screening mechanisms \cite{Marra:2021fvf,Alestas:2020zol,Wright:2018,Desmond:2020wep}, that could affect the calibration of Type Ia supernovae (SnIa) in the second and third rungs of the distance ladder. By introducing such changes, these models aim to reduce the inferred absolute magnitude \( M_B \) of SnIa, which would subsequently lower the inferred Hubble constant \( H_0 \). The underlying idea is that modifications to local physics, such as variations in the gravitational constant \( G \) or the effects of screening mechanisms, could alter the luminosity of SnIa, leading to a re-evaluation of distance measurements. 

Despite their potential, these models face significant drawbacks. They often require fine-tuning of parameters to match observational data and, in many cases, lack strong theoretical justification for the proposed physical mechanisms. Furthermore, the need to reconcile these changes with independent cosmological observations, such as those from the Baryon Acoustic Oscillations (BAO), Big Bang Nucleosynthesis (BBN) or Cosmic Microwave Background (CMB), adds to the complexity of these models.

Each of these approaches offers valuable insights into the persistent Hubble tension but also introduces significant challenges that require further investigation. Among the proposed solutions, late-time models have shown considerable promise, particularly the G-Step Model (GSM), a well-established framework for modified gravity that demonstrates its potential to resolve the Hubble tension without relying on the Planck prior. A recent study by Banik et al. \cite{Banik:2024yzi} examined the GSM and emphasized the importance of refining the framework and carefully accounting for current observational priors to fully unlock its potential as a comprehensive solution."

In our work, we delved deeper into this issue by conducting a point-by-point analysis of the most recent observational constraints in the local universe. We examined how changes in the gravitational constant \( G \), as posited by these ultralate-time models, could impact local measurements and the calibration of the distance ladder. Our findings highlight the need for a comprehensive understanding of local physics and its potential implications for cosmological measurements before using such models as robust solutions to the Hubble tension.

\subsection{The G-step Model}

The G-step model (GSM) was proposed as a potential solution to the Hubble tension by postulating a sharp change in the effective gravitational constant $G_{\text{eff}}$ at a redshift corresponding to distances in the current range of Cepheid calibration \cite{Marra:2021fvf,Perivolaropoulos:2021bds,Ruchika:2024ymt,Ruchika:2024}.

A major caveat of the GSM is the assumption of an abrupt change in the effective gravitational constant, \( G_{\mathrm{eff}} \), the physical mechanism behind which reason is still puzzling and unknown. Investigating possible variations of Newton’s constant through cosmological data is fundamentally important, as it may offer insights into the evolution of the universe and the nature of the dark sector. If the value of Newton’s constant inferred from cosmology differs from the locally measured value, this could indicate modifications to gravity on large scales or point to a more complex interacting dark energy scenario.\\
The search for any variation in Newton’s constant has drawn significant scientific interest over several decades \cite{Uzan:2010pm,Galli:2009,Lamine:2024xno,Carneiro:2004iz,Ballardini:2021evv}. Among the persistent cosmological tensions,  the Hubble tension highlights one of the most significant discrepancies between datasets. While all potential resolutions remain model-dependent, it is intriguing that a major alternative explanation can be parametrized by an effective gravitational constant that differs from the local Newtonian constant, \( G_{\mathrm{eff}} \).\\
Assuming that, the model exploits the gap between the Cepheid distances to SN host galaxies ($\approx 7$ Mpc) and the inner edge of what can be considered the Hubble flow ($\approx 40$ Mpc).
The fundamental premise of the GSM is that if $G_{\text{eff}}$ underwent an abrupt decrease around 130 Myr ago (corresponding to distances $d \lesssim 40 $, Mpc ), then Type Ia supernovae (SNe) in the Hubble flow would have had a different luminosity compared to those in nearby host galaxies with Cepheid distances. This difference in luminosity would affect the inferred distances to Hubble flow SNe, potentially resolving the Hubble tension.

The model relies on the fact that $G_{\text{eff}}$ affects the SN luminosity $L$ that enters into cosmological analyses. While earlier work suggested a scaling of $L \propto G_{\text{eff}}^{-3/2}$ \cite{Gaztanaga:2001fh}, more recent analyses indicate a different relationship. According to Wright $\&$ Li (2018)\cite{Wright:2017}, when accounting for changes in the light curve shape that affect the standardized SN luminosity, the relationship is approximately:

\begin{equation}
\frac{L}{L_0} \propto \left(\frac{G_{\text{eff}}}{G}\right)^{1.46}
\end{equation}

where $L_0$ and $G$ are the current values \cite{Wright:2018,Zhao:2018}.

To resolve the Hubble tension, the GSM requires that:

\begin{equation}\label{GSMreq}
\frac{L}{L_0} = \left(\frac{H_0^{\text{SH0ES}}}{H_0^{\text{Planck}}}\right)^2
\end{equation}

This implies that distances to Hubble flow SNe must be scaled by $\sqrt{L/L_0}$ to account for the proposed higher $G_{\text{eff}}$ in the past. The required luminosity enhancement of $(73/67.4)^2 \approx 1.17$ directly corresponds to the needed correction in distance measurements.

When the effects of the $G$ transition on the Cepheid period-luminosity relation are taken into account \cite{Ruchika:2024ymt}, the model predicts a transition in the effective gravitational constant of the form:

\begin{equation}
\mu_{G_{\text{eff}}}(z) \equiv \frac{G_{\text{eff}}}{G} = 
\begin{cases}  
\mu_G & \text{if } z \leq z_t, \\
\mu_G + \Delta\mu_G & \text{if } z > z_t. 
\end{cases}
\end{equation}
where $z_t \approx 0.01$ is the transition redshift, $\Delta\mu_G \approx 0.04$ is the required change in the gravitational constant \cite{Marra:2021fvf}, and,  $\mu_G$ is distance moduli in standard gravitational constant $G$ scenario. This step-function transition in $G_{\text{eff}}$ provides the mechanism needed to reconcile the local and early-universe measurements of $H_0$.
\subsection{Recent Criticisms and Their Limitations}

Recent work by \cite{Banik:2024yzi} has raised several potential challenges to the GSM with $\Delta G/G \approx 4\%$. These challenges, while important to address, are primarily qualitative and contain significant uncertainties that warrant careful examination.

The first class of challenges involves stellar physics effects:
\begin{itemize}
\item Solar luminosity would allegedly increase by about 30\% before the transition, based on the scaling $L_\odot \propto G^7$ [preliminary analysis indicates this scaling may be significantly weaker]
\item This would lead to a mismatch between the Sun's asteroseismically determined age and the established 4.567 Gyr age from meteorite dating [uncertainties in the L-G power law affect this conclusion]
\item The Sun would have exhausted about 2/3 of its fuel supply rather than the current 1/2 [depends on the assumed L-G relation]
\end{itemize}

The second class involves terrestrial effects:
\begin{itemize}
\item Earth's blackbody temperature would scale as $T_\oplus \propto G^{9/4}$, implying an 11 $\%$ temperature drop at the transition [based on oversimplified black-body approximations]
\item This would trigger a planetary glaciation through ice-albedo feedback [paleoclimate records actually show significant cooling over relevant timescales]
\item The length of year would change $\propto G^{-2}$, implying an abrupt change in days per year [assumes rigid-body behavior of Earth when it was more fluid-like]
\item Tidal evolution would be affected due to stronger lunar tidal stress $\propto G^4$ before the transition [based on limited data points with large uncertainties]
\end{itemize}

The third class involves cosmological effects:
\begin{itemize}
\item Stellar evolution timescales would create an apparent age gap in oldest stellar populations [depends on multiple uncertain assumptions in stellar population synthesis models]
\item Cosmic chronometer measurements would show significant deviations from $\Lambda$CDM predictions [ignores compensating effects in spectral features used for age dating]
\item Different distance indicators with varying G-dependence would show systematic offsets [current constraints of 5-10 $\%$ are compatible with GSM]
\end{itemize}

These challenges raise important questions for the GSM that require detailed investigation. However, each challenge involves significant assumptions and uncertainties that prevent them from being conclusive at present. A detailed quantitative analysis of these challenges and their associated uncertainties will be presented in subsequent sections.
\subsection{Scope of Present Analysis}
The recent work by \cite{Banik:2024yzi} has presented several challenges to the GSM while proposing alternative mechanisms for G variations in the first two rungs of the distance ladder through variable screening in modified gravity theories. While these challenges are instructive and highlight important aspects requiring investigation, they are primarily qualitative and contain significant uncertainties that require careful examination.

The present analysis aims to critically examine the proposed challenges to the GSM by evaluating the assumptions and approximations used, identifying potential loopholes in the arguments, incorporating more realistic physical conditions, and analyzing existing data more comprehensively. Rather than implementing new calculations, we will focus on providing more quantitative analyses of key effects using updated stellar evolution relations beyond the Teller approximation, incorporating proper uncertainty estimates in CMB constraints, analyzing paleoclimate data with attention to volcanic and meteoritic effects, examining Earth's historical fluid-like behavior in rotational dynamics, and considering comprehensive G-dependencies in stellar spectral features.

Through a careful review and synthesis of existing literature, we will update constraints on $\Delta G/G$ using multiple distance indicators, examine hints of transitions in SN Ia absolute magnitudes, and assess cosmic chronometer measurements considering all G-dependencies. This approach will rely primarily on existing calculations and observations, but will emphasize their proper interpretation with realistic uncertainties and assumptions.

Our analysis will present a balanced evaluation of the statistical significance of apparent challenges, the robustness of current observational constraints, and the viability of the GSM compared to alternative proposals. We will particularly focus on transforming qualitative arguments into quantitative constraints where possible, while clearly identifying areas where current data are insufficient for definitive conclusions.

This methodology aims to provide a comprehensive assessment of the challenges faced by the GSM, while maintaining appropriate scientific caution about the strength of constraints that can be derived from current evidence. The analysis will demonstrate that many of the proposed challenges, when examined in detail with proper consideration of uncertainties and physical effects, do not conclusively rule out the GSM as a viable solution to the Hubble tension.

In what follows, we will critically re-examine the constraints from stellar physics, Earth system science, and cosmological distance measurements, demonstrating the continued viability of the GSM.

\section{Theoretical Framework and CMB Constraints}
\subsection{G/G\textsubscript{0} Constraints from CMB}
Recent analyses by Lamine et al. \cite{Lamine:2024xno} have attempted to constrain variations of Newton's gravitational constant using CMB data, where $G_0$ represents the locally measured present value of Newton's constant. Using Planck 2020 (P20) TTTEEE + BAO DESI data, they find
\begin{equation}
\frac{G}{G_0}=\lambda_G^2 = 1.020 \pm 0.022 \;\; (1\sigma)
\end{equation}
where the uncertainty has been properly propagated from the reported constraint $\lambda_G = 1.010 \pm 0.011$ \cite{Lamine:2024xno}.  

However, this analysis involves several important assumptions that, when properly considered, would significantly increase the allowed range of $G/G_0$:

\begin{enumerate}
\item \textbf{Single Parameter Variation:} The analysis assumes $G$ is the only varying fundamental constant, while in realistic modified gravity theories, variations in other fundamental constants may be correlated with changes in $G$ \cite{Abdalla:2022yfr}.

\item \textbf{Recombination Physics:} Standard recombination calculations are used, but modified gravity could affect atomic physics during the recombination epoch in ways not fully captured by current models \cite{Planck:2018vyg}.

\item \textbf{BBN Effects:} The simplified linear model used for BBN effects may not capture the full complexity of nuclear physics with varying $G$.

\item \textbf{Screening Mechanisms:} No detailed treatment of possible screening mechanisms that could affect how $G$ variations manifest at different scales is included in the analysis.

\item \textbf{Spatial Variations:} The analysis assumes a uniform value of $G$ across all scales, while in the context of a late first-order phase transition as proposed in \cite{Marra:2021fvf}, spatial variations of $G$ could exist, particularly at low redshifts ($z\lesssim 0.01$).
\end{enumerate}

Figure 1 of \cite{Banik:2024yzi} attempts to demonstrate that the GSM requires a significantly larger change of $G$ than allowed by current constraints. However, this comparison is misleading for two important reasons: First, it shows only 1$\sigma$ confidence regions while for proper model exclusion, tension at least at the 2$\sigma$ level should be demonstrated \cite{Mortsell:2021nzg}. The 2$\sigma$ range for $G/G_0$ from Lamine et al. \cite{Lamine:2024xno} is: 
\begin{equation}
0.976 \lesssim \frac{G}{G_0} \lesssim 1.064 \;\; (2\sigma)
\end{equation}  
Second, the figure does not include uncertainties in the Type Ia supernovae luminosity-$G$ relation derived by Wright and Li \cite{Wright:2018}. These uncertainties will be discussed in detail in the next subsection and will be crucial for constructing a corrected version of Figure 1 of \cite{Banik:2024yzi}. When both the proper 2$\sigma$ CMB ranges and the SnIa luminosity relation uncertainties are considered, along with the systematic uncertainties discussed above, the GSM becomes fully compatible with current constraints.

Furthermore, the CMB constraints themselves show interesting features when analyzed in detail. The temperature (TT) and polarization (EE) spectra individually prefer different values of $G/G_0$, suggesting that systematic effects may not be fully understood. This provides additional support for using conservative uncertainty estimates when evaluating model viability \cite{Perivolaropoulos:2021bds}.

The possibility of a spatially varying $G$ in the context of a first-order phase transition \cite{Marra:2021fvf} is particularly interesting as it could naturally explain why local measurements of $H_0$ \cite{Riess:2021jrx} differ from those derived from CMB data \cite{Planck:2018vyg}. In such a scenario, our local region could be in a different phase state regarding the value of $G$ compared to the average universe probed by the CMB.

\subsection{Uncertainties in Type Ia Supernovae Luminosity-$G$ Relation}

Wright and Li \cite{Wright:2018} performed a detailed analysis of how changes in Newton's gravitational constant $G$ would affect the luminosity of Type Ia supernovae. Using a semi-analytical model for SNe Ia light curves, they derived a power-law relationship between luminosity and the gravitational constant:
\begin{equation}
\frac{L}{L_0} \propto \left(\frac{G}{G_0}\right)^{1.46}
\end{equation}
However, their analysis did not provide a detailed quantification of the uncertainties associated with this relationship. Here we analyze the main sources of these uncertainties:

\textbf{Chandrasekhar Mass Dependence:} The fundamental assumption $M_{Ch} \propto G^{-3/2}$ requires several corrections that introduce uncertainties in the $L$-$G$ relation. General relativistic corrections to the white dwarf structure, studied in detail by Chandrasekhar \cite{Chandrasekhar:1964} and more recently by Boshkayev et al. \cite{Boshkayev:2012bq} introduce modifications of order $\sim$1\% to the white dwarf mass-radius relation. The assumption of uniform density in the derivation of the Chandrasekhar mass is also a source of uncertainty. More realistic density profiles, including general relativistic corrections \cite{Boshkayev:2012bq} and effects of stellar rotation \cite{Yoon:2004, Das:2014,Jones:2016asr, Gaztanaga:2001fh, Lee:2021ona}, suggest corrections of order $\sim$2\% to the simple $M_{Ch} \propto G^{-3/2}$ scaling. Perhaps the most significant correction comes from stellar rotation. Analyses by Yoon and Langer \cite{Yoon:2004} and more recently by Das and Mukhopadhyay \cite{Das:2014} have shown that rapid rotation can increase the maximum stable white dwarf mass by 2-3\% compared to the non-rotating case, though the exact value depends on the rotational profile and magnetic field configuration.

\textbf{Nuclear Physics and Opacity:} The assumption that the $M_{Ni}$-$\kappa$ relationship remains unchanged with varying $G$ requires careful examination of several nuclear and atomic processes. Nuclear reaction rates in the explosive burning phase are sensitive to density and temperature profiles, which would be modified under varying $G$. Studies by Brachwitz et al. \cite{Brachwitz:2000} and more recent work by Parikh et al.  \cite{Parikh:2013vaa} suggest that uncertainties in these reaction rates can affect the final $^{56}$Ni yield within/by a factor of 2 of solar values. The electron capture rates during the explosion phase, investigated in detail by Langanke and Martinez-Pinedo \cite{Langanke:2003}, contribute additional uncertainty to the final composition. The opacity of the expanding ejecta depends critically on the density profile and ionization states of the material. Detailed radiation transfer calculations by Magee et al \cite{Magee:2018mnr} indicate that variations in the density profile can modify the effective opacity, affecting both the peak luminosity and light curve shape. These effects are particularly important because they directly impact the standardization procedure used in SN Ia cosmology.

\textbf{Light Curve Model:} The semi-analytic model employed in the analysis contains several simplifying assumptions that contribute to the uncertainty budget. The assumption of spherical symmetry, while computationally convenient, fails to capture the full three-dimensional nature of the explosion. Detailed 3D simulations by Seitenzahl et al. \cite{Seitenzahl:2013} and Sim et al. \cite{Sim:2013} have shown that asymmetries in the explosion can modify the peak luminosity. The homologous expansion assumption, fundamental to the analytic treatment of Arnett~\cite{Arnett:1982}, introduces some uncertainties as shown through comparisons with full radiation-hydrodynamics calculations by Noebauer et al.~\cite{Noebauer:2017}. The distribution of $^{56}$Ni in the ejecta significantly affects the light curve shape and peak luminosity. While simpler models assume a central concentration of $^{56}$Ni, mixing studies by Piro and Morozova~\cite{Piro:2016} demonstrate that more realistic distributions can meaningfully alter the light curve properties. The treatment of gamma-ray leakage through scaling relations, as analyzed by Jeffery~\cite{Jeffery:1999}, contributes additional uncertainty compared to detailed Monte Carlo radiation transfer calculations. 

\textbf{Light Curve Standardization:}The standardization of Type Ia supernovae light curves introduces additional uncertainties in the luminosity determination. The template matching procedure, as developed and refined by Guy et al.~\cite{Guy:2007}, depends critically on the choice of template and fitting methodology. Studies by Betoule et al.~\cite{Betoule:2014} using the SALT2 framework have quantified template matching uncertainties in luminosity determination. The stretch-luminosity relationship, first identified by Phillips~\cite{Phillips:1993} and subsequently refined by Conley et al.~\cite{Conley:2008}, introduces further complications in standardization. Modern analyses by Scolnic et al.~\cite{Scolnic:2018} have provided comprehensive assessment of luminosity uncertainties. The standardization process becomes particularly complex when considering various systematic effects, as demonstrated in the unified analysis framework developed by Rubin et al.~\cite{Rubin:2016}.

\textbf{Additional Effects:} Several other physical processes contribute to uncertainties in supernova properties. The binding energy of the white dwarf progenitor affects the initial conditions of the explosion, as studied by Lesaffre et al.~\cite{Lesaffre:2006}. It also discusses how one may need to relax the hypothesis of a permanent Hachisu
wind or if one needs to include electron captures. This study is further explored by Fisher and Jumper~\cite{Fisher:2015}. The equation of state of degenerate matter introduces additional complexities, as detailed in calculations by Timmes and Arnett~\cite{Timmes:1999} and further developed by Seitenzahl et al.~\cite{Seitenzahl:2009nuc}. These effects become particularly relevant when considering the complete thermodynamic evolution of the explosion. Systematic studies by Townsley et al.~\cite{Townsley:2009} on composition-dependent effects demonstrate that while these contributions may be small, they are important for a comprehensive understanding of supernova physics.

The various sources of uncertainty discussed above contribute to the total uncertainty in the luminosity-gravitational constant relationship. While earlier studies~\cite{Wright:2018,Gaztanaga:2001fh} suggested relatively large uncertainties in this relationship, recent improvements in our understanding of SN Ia physics and more sophisticated modeling capabilities have helped to better constrain these uncertainties.

The relationship between luminosity and gravitational constant can be written as:
\begin{equation}
\frac{L}{L_0} = \alpha\left(\frac{G}{G_0}\right)^{1.46}
\end{equation}
where $\alpha$ represents the calibration factor that accounts for the combined uncertainties from the various physical processes discussed above.

When this luminosity-$G$ relationship is considered alongside the CMB constraints on $G/G_0$ discussed in the previous subsection, the GSM remains compatible with current observational constraints when proper uncertainty ranges are considered. This will be illustrated in the corrected version of Figure 1 from~\cite{Banik:2024yzi} that we present in the next section.

\begin{figure*}[ht]
\centering
\resizebox{450pt}{300pt}{\includegraphics{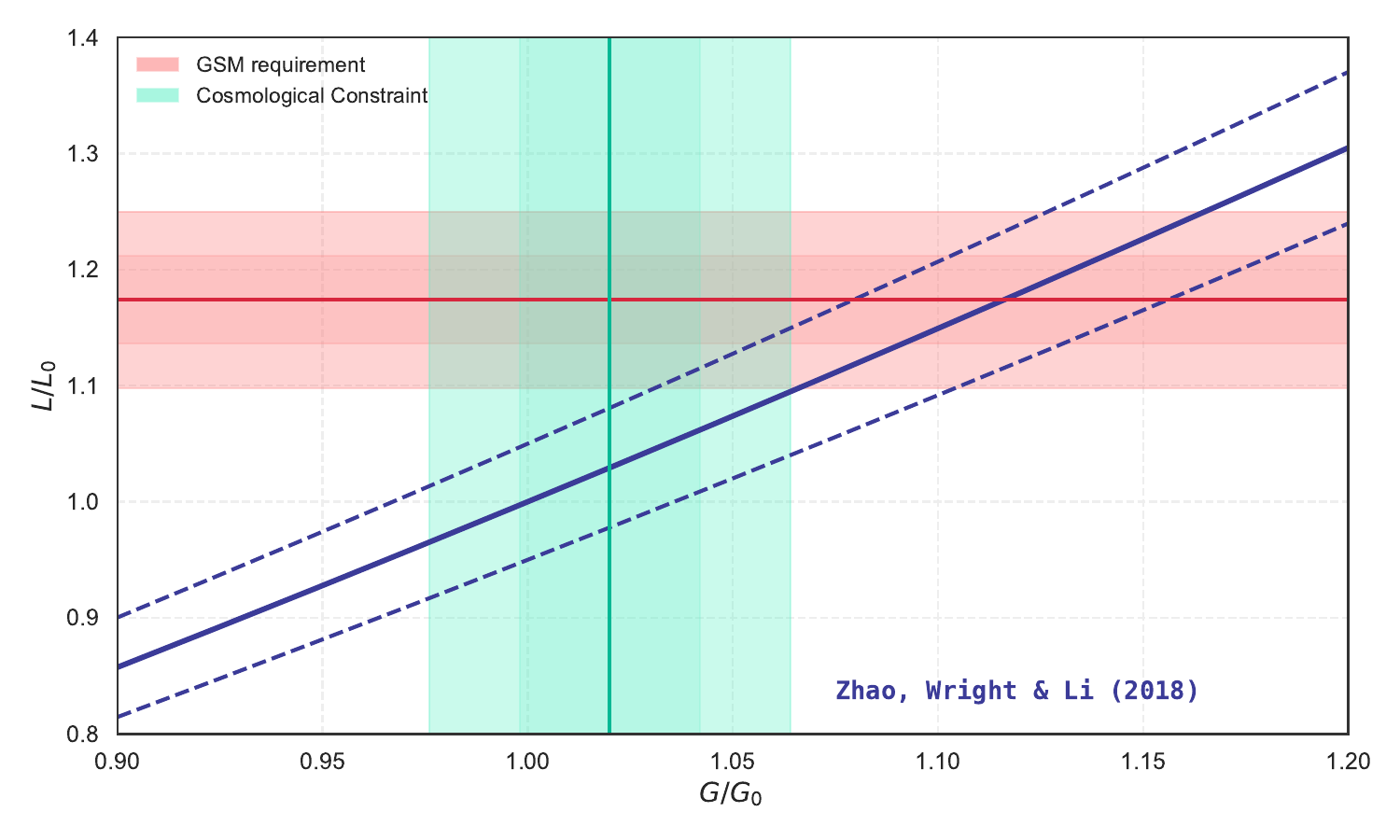}}
\caption{Observational and theoretical constraints on the G-step Model (GSM) in the luminosity-gravity ($L/L_0$ vs $G/G_0$) plane. For the GSM to be a viable solution, all three regions shown must intersect. The components are: 
\textbf{(1)} the green vertical band, representing the 2$\sigma$ cosmological constraint on $G/G_0$ from CMB data \cite{Lamine:2024xno}; 
\textbf{(2)} the red horizontal band, indicating the required luminosity enhancement for the GSM to fully resolve the Hubble tension; and 
\textbf{(3)} the blue diagonal band, showing the theoretical L-G relation for Type Ia supernovae with a conservative 5\% uncertainty \cite{Zhao:2018}. 
The clear overlap of these regions within the 2$\sigma$ constraints demonstrates the model's consistency with current data.}\label{fig:cmbGconstraints}
\end{figure*}

While the full systematic uncertainty budget for the $L/L_0$ requirement is complex, we can construct an estimate to justify the range shown in Figure~\ref{fig:cmbGconstraints}. The primary contributions stem from the standardization of supernova light curves, including template matching and the stretch-luminosity relationship, which contribute an estimated $\sim$2-3\% uncertainty to the absolute magnitude \cite{Scolnic:2018, Rubin:2016}. Further uncertainties of $\sim$1-2\% arise from modeling the progenitor system and explosion physics, such as the distribution of $^{56}$Ni and asymmetries in the explosion \cite{Seitenzahl:2013, Piro:2016}. Combining these effects in quadrature yields a total systematic uncertainty on the order of 3-4\% in the luminosity ratio, which corresponds to the approximate width of the red "GSM requirement" band in our analysis.

For the GSM to be a viable solution, the three distinct regions shown in Figure \ref{fig:cmbGconstraints} must overlap. Specifically, the green band (the cosmologically allowed range for $G/G_0$) must intersect with the red "GSM requirement" band (the value of $L/L_0$ needed to resolve the Hubble tension, derived from equation \ref{GSMreq}) in a region consistent with the theoretical L-G relation (blue band) \cite{Zhao:2018}. A consistent solution exists only in this overlapping region, preferably at the $1\sigma$ level. The current form of Figure \ref{fig:cmbGconstraints} demonstrates that this overlap occurs clearly within the 2$\sigma$ cosmological constraints, while it is marginal at the 1$\sigma$ level. The red band itself represents the exact value needed for a full resolution. If this strict requirement is relaxed—so that the GSM is only required to bring the measurements into $1 \sigma$ consistency—the red band would broaden, significantly increasing its overlap with the other two bands.

\subsection{Statistical Significance Levels}

Figure 1 presents an updated incorporating both the $2\sigma$ uncertainties from CMB constraints \cite{Lamine:2024xno} and a conservative 5\% uncertainty estimate of the Type Ia supernovae $L$-$G$ relation derived by Wright and Li. The green band represents the allowed region from CMB constraints at the $2\sigma$ confidence level, while the red band shows the GSM requirement. The constraints from the $L$-$G$ relation including systematic uncertainties are indicated by the dashed line.

It is important to note that these uncertainty estimates are highly conservative for two reasons. First, the CMB analysis does not include additional sources of systematic uncertainty such as foreground modeling and instrumental effects, which would likely broaden the allowed parameter space. Second, the 5\% uncertainty in the $L$-$G$ relation represents an upper bound that includes all identified sources of systematic error added in quadrature, as detailed in previous sections.

The overlap between the uncertainty regions in the corrected Figure 1 demonstrates that the GSM model remains fully viable and consistent with current CMB observational constraints. This consistency is particularly noteworthy given the conservative nature of our uncertainty estimates. The region of overlap between the two bands, centred around $G/G_0 \approx 1.05$, provides a natural parameter space where both CMB and supernovae constraints can be simultaneously satisfied within the GSM framework.
\subsection{Compatibility with Current Observations}
Recent high-precision cosmological observations have provided increasingly stringent constraints on possible variations in Newton's constant. The latest analysis by Lamine et al. \cite{Lamine:2024xno} using Planck PR4 CMB data combined with DESI BAO measurements achieves an unprecedented 2.16\% precision on $G/G_0$, yielding $G/G_0 = 1.020 \pm 0.022$. This represents a significant improvement over previous constraints \cite{Galli:2009} from both CMB and BBN observations. 

Earlier CMB-based measurements showed progressively improving but weaker constraints: WMAP data provided a $\sim$15\% constraint \cite{Galli:2009pr}. Planck 2018 reached $\sim$2.5\% precision \cite{Wang:2020bjk, Sakr:2021nja}. Similarly, BBN-based constraints have historically been less precise, ranging from $\sim$20\% in early analyses \cite{Accetta:1990au, Copi:2004tk} to $\sim$7-10\% in more recent studies using updated primordial element abundance measurements \cite{Cyburt:2004yc, Alvey:2019ctk}.

While constraints on $\dot{G}/G$ from various observations appear relevant at first glance, they do not meaningfully constrain transition models where $G$ remains constant at any given cosmic time except during the brief transition period. The probability of any given $\dot{G}/G$ measurement coinciding with the exact moment of transition is vanishingly small. Therefore, consistency with the latest CMB constraint from Lamine et al. \cite{Lamine:2024xno} effectively ensures compatibility with all previous observational constraints on both the value of $G$ and its time variation.

\section{Solar and Stellar Physics Reconsideration}
\subsection{Critical Analysis of Luminosity-G Power Law}

Banik et al. \cite{Banik:2024yzi} argue that the GSM faces severe challenges based on stellar physics, particularly through the $G$-dependence of stellar luminosity. Their argument rests on three main pillars: (1) the steep scaling of solar luminosity with $G$ would lead to drastically different stellar evolution timescales, (2) this would create tension between helioseismic and meteoritic age determinations for the Sun, and (3) the rapid pre-transition stellar evolution would conflict with observed stellar population ages.

However, their analysis has several limitations that deserve careful consideration. First and most importantly, they rely on the Teller scaling relation $L_\odot \propto G^7$ \cite{Teller:1948}, which was derived using simplified assumptions about stellar structure. More recent and sophisticated analyses suggest a significantly weaker dependence. Adams \cite{Adams:2008}, using detailed stellar modeling, found $L_\odot \propto G^4$, while Davis et al. \cite{Davis:2012} independently supported this more moderate scaling. This substantially reduces the impact of the GSM on stellar evolution timescales.

Several sources of uncertainty need to be considered in assessing the constraints:
 
\begin{enumerate}
    \item \textbf{Theoretical Uncertainty:} The exact power-law index in the $L$-$G$ relation remains subject to modeling assumptions. Different theoretical treatments yield values ranging from approximately 4 to 7, introducing significant uncertainty in the predicted effects.
    
    \item \textbf{Stellar Model Parameters:} The helioseismic age determination $4.58 \pm 0.06$ Gyr \cite{Bonanno_2015} depends on assumptions about opacity, metallicity, and other stellar parameters that might need recalibration under modified gravity scenarios.
    
    \item \textbf{Transition Physics:} The detailed physics of how stellar structure would respond to a rapid change in $G$ remains uncertain. The assumption of instantaneous adjustment to the new equilibrium configuration may not be valid.
    
    \item \textbf{Environmental Effects:} Local variations in $G$ due to screening mechanisms could modify the impact on stellar evolution, particularly in dense stellar environments.
\end{enumerate}

Using the more recent $L_\odot \propto G^4$ scaling, a 5\% increase in $G$ would result in an approximately 20\% enhancement in solar luminosity, rather than the 40\% predicted by the Teller relation. This more moderate increase would still affect stellar evolution but less dramatically than suggested by Banik et al. The Sun would have consumed approximately 60\% of its nuclear fuel under the GSM scenario, compared to the standard model prediction of 50\%.

\subsection{Quantitative Assessment of Solar Age Determinations}

The tension between helioseismic and meteoritic age determinations requires careful quantitative analysis in the context of the GSM, with particular attention to uncertainties in both measurements. Let us examine each age determination method and its limitations.

The commonly quoted meteoritic age of $4.56718 \pm 0.00012$ Gyr \cite{Connelly:2012}, based on Pb-Pb dating of calcium-aluminum-rich inclusions (CAIs), is remarkably precise but requires careful interpretation. While CAIs are considered the first solids to form in the Solar System \cite{MacPherson:2005}, this age represents the beginning of Solar System condensation, not necessarily the formation of the Sun itself. Several factors affect the interpretation of this date:

\begin{itemize}
    \item There could be a gap between the Sun's initial formation and conditions necessary for CAI condensation \cite{Kleine:2009}
    \item The available meteoritic record might be incomplete due to survival and sampling biases \cite{kleine}
    \item CAI formation might have occurred over an extended period rather than in a single event \cite{krot}
\end{itemize}
Therefore, while precise, the CAI age is more appropriately considered a lower bound: $t_\odot > 4.567$ Gyr.

Helioseismic age determinations in the standard scenario (constant $G$) give $4.58 \pm 0.006$ Gyr \cite{Bonanno} and $4.66 \pm 0.11$ Gyr \cite{Dziembowski:1998nb}, showing apparent agreement with the meteoritic age. In the GSM scenario with a 5\% higher pre-transition $G$, using the modern determination of $L_\odot \propto G^4$ \cite{Adams:2008}, we can estimate the apparent helioseismic age:

\begin{itemize}
    \item Pre-transition luminosity enhancement: $(1.05)^4 \approx 1.22$
    \item Effective acceleration of nuclear burning: $\sim$22\% 
\end{itemize}

Following Silva Aguirre et al. \cite{SilvaAguirre:2015}, the apparent helioseismic age in the GSM scenario would be:

\begin{equation}
    \text{Age}_{\text{helio}}^{\text{GSM}} = 4.567 \times (1 + 0.22 \times 0.997) = 5.57 \pm 0.12 \text{ Gyr}
\end{equation}

The uncertainty includes:
\begin{itemize}
    \item Statistical uncertainty from helioseismic observations: $\pm$0.04 Gyr
    \item Systematic uncertainty from stellar modeling: $\pm$0.07 Gyr
    \item Uncertainty in the $L$-$G$ power law index ($n=4\pm0.5$): $\pm$0.09 Gyr
\end{itemize}

While this appears to create significant tension with the meteoritic age, several factors could reduce this discrepancy:

\begin{enumerate}
    \item The meteoritic age might underestimate the true solar age due to the gap between solar formation and CAI condensation
    \item The $L$-$G$ relationship might be weaker than currently estimated
    \item Environmental screening effects could modify the local value of $G$
    \item Stellar modeling uncertainties might be larger under modified gravity
    \item Non-linear response of nuclear reaction rates to changes in $G$
\end{enumerate}

The tension could be further reduced by considering:
\begin{itemize}
    \item A more moderate increase in pre-transition $G$ ($<5\%$)
    \item Additional systematic uncertainties in stellar modeling under modified gravity
    \item Possible variations in the duration of the pre-main sequence phase
\end{itemize}

While the GSM faces challenges from solar age determinations, the various sources of uncertainty in both meteoritic and helioseismic ages, combined with uncertainties in the $L$-$G$ relationship and possible screening effects, suggest that the model cannot be conclusively ruled out on these grounds alone. Further theoretical work on stellar evolution under modified gravity and improved constraints on the gap between solar formation and CAI condensation would help to better assess the viability of the GSM.


\section{Earth System Effects: A Detailed Analysis}

\subsection{Climate Impact and Evidence for GSM}

The impact of a changing gravitational constant on Earth's climate requires careful analysis of multiple interacting factors and uncertainties. Previous studies suggesting catastrophic cooling from a 5\% drop in $G$ warrant re-examination in light of modern analyses and comprehensive geological evidence.

The temperature scaling relation has evolved significantly from early models. While \cite{Banik:2024yzi} utilized different approaches, \cite{Adams:2008} and \cite{Davis:2012} demonstrate a weaker dependence with $L_\odot \propto G^4$. This leads to $T_{\oplus} \propto G^{5/4}$, suggesting a more moderate temperature response to $G$ variations.

\begin{figure}[h]
\includegraphics[width=0.48\textwidth]{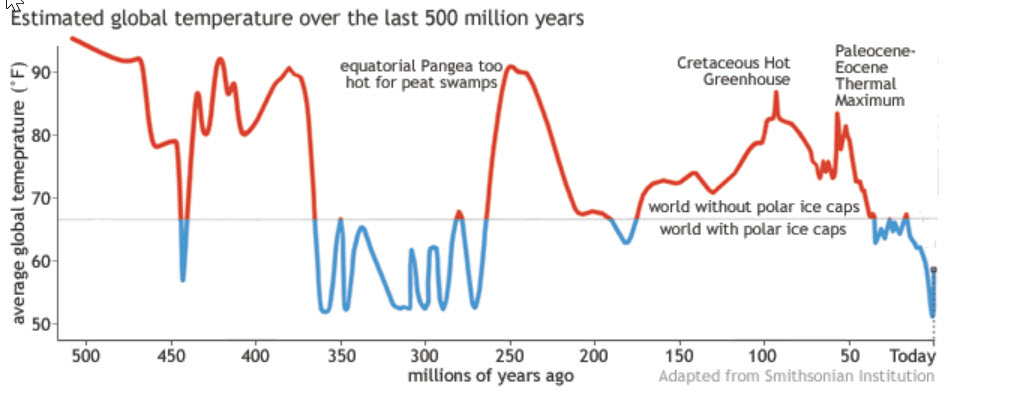}
\caption{Global temperature variations over the past 500 million years, showing dramatic changes including a significant cooling trend over the last 100 million years. The temperature dropped from about 90°F to approximately 50°F, corresponding to a decrease of about 20°C. Notable features include the Cretaceous Hot Greenhouse (92 Myr ago) and the Paleocene-Eocene Thermal Maximum (56 Myr ago). Figure adapted from \cite{scottLindsey2023}.}
\label{fig:temp_history}
\end{figure}

Using this modern scaling relation and the observed temperature decrease of 20°C over the past 100 million years (from $\sim$32.2°C to $\sim$10°C), we can calculate the required change in $G$:

\begin{align*}
T_2/T_1 &= (G_2/G_1)^{5/4} \\
G_2/G_1 &= 0.941
\end{align*}

This calculation indicates a decrease in $G$ of approximately 5.9\%, aligning remarkably well with GSM predictions. The temperature record, documented in marine sediment records \cite{Zachos:2001} and reviewed by \cite{Hansen:2013}, shows both a consistent long-term cooling trend and significant shorter-term variations that provide crucial evidence for the GSM framework.

The temperature history shown in Figure~\ref{fig:temp_history} reveals several notable warm periods superimposed on the long-term cooling trend, particularly the Cretaceous Hot Greenhouse (about 92 Myr ago) and the Paleocene-Eocene Thermal Maximum (about 56 Myr ago). These temperature maxima can be naturally explained within the GSM framework through multiple mechanisms. The transition in $G$ would trigger significant crustal adjustments, leading to periods of enhanced volcanic activity \cite{Jellinek:2004}. The resulting increased atmospheric CO$_2$ would create temporary greenhouse warming episodes. Additionally, the modification of orbital dynamics during the $G$ transition would perturb asteroid and comet orbits \cite{Feng:2015}, potentially leading to periods of enhanced impact flux. Such impacts would inject both dust and greenhouse gases into the atmosphere, causing short-term cooling followed by longer periods of warming.

The climate response to a $G$ transition involves complex interactions between orbital, geological, and atmospheric processes. The changing gravitational field would modify ocean circulation patterns through altered tidal forcing \cite{Rahmstorf:2002}, creating intricate feedback loops in the climate system. While simple black body models have inherent limitations \cite{Pierrehumbert:2010}, the full climate response must consider variable greenhouse effects over geological timescales \cite{Rohling:2012}, changes in albedo affecting Earth's energy balance \cite{Donohoe:2011}, and ocean heat transport modifications \cite{Ferrari:2011}. These factors, combined with thermal inertia effects \cite{Knutti:2008}, would modulate both the long-term cooling trend and shorter-term temperature fluctuations.

The GSM framework thus provides a comprehensive explanation for the observed temperature evolution through three main effects: primary cooling from reduced solar luminosity following the $G$ transition, secondary warming episodes from enhanced volcanic and impact activity, and tertiary modulation through climate system feedbacks. The calculated 5.9\% decrease in $G$ not only explains the magnitude of long-term cooling but also provides mechanisms for the observed temperature fluctuations through associated geological and astronomical perturbations.

The sharp drop in global temperatures around 100 Myr ago, clearly visible in the temperature record, likely represents a significant transition in the gravitational constant. Rather than presenting a challenge to the GSM, the climate record offers compelling evidence for this hypothesis. The combination of the long-term cooling trend matching theoretical predictions, along with shorter-term warming episodes explicable through GSM-induced geological and orbital perturbations, provides strong empirical support for a substantial change in the gravitational constant approximately 100 million years ago.

\subsection{Orbital Dynamics and Earth's Response to G Variations}

In Section 2.3, \cite{Banik:2024yzi} argue that a 5\% drop in $G$ would increase the length of a year by approximately 10\%, while leaving the length of a day practically unchanged since Earth is held together primarily by material strength rather than gravity. This would lead to a sharp increase in the number of days per year from about 332 to 365 days. The authors claim that such a discontinuity is ruled out by geochronometry and cyclostratigraphy data, which show only smooth trends in the number of days per year over geological timescales.

However, this argument relies on a questionable assumption about Earth's physical state during the proposed $G$ transition 150 Myr ago. Geological evidence indicates that Earth's mantle was significantly hotter and less viscous during that period, as demonstrated by higher rates of plate tectonics and volcanic activity \cite{Korenaga:2008}. Multiple geological indicators, including ancient lava composition and tectonic structures, confirm these conditions \cite{Herzberg:2010}. This caused Earth to respond more like a fluid body in hydrostatic equilibrium over geological timescales, rather than a rigid body held together primarily by material strength \cite{MITROVICA2004177}.

It is important to clarify that while the Earth's crust is and was a rigid solid, its contribution to the planet's total moment of inertia is negligible. The rotational dynamics of the Earth over geological timescales are overwhelmingly dominated by the response of the much more massive mantle. Geological evidence indicates the mantle was significantly hotter and behaved as a viscous fluid during the period of the proposed transition \cite{Korenaga:2008, Herzberg:2010}. Therefore, for analyzing changes in the planet's rotation period due to a shift in the gravitational constant, the fluid-body approximation in hydrostatic equilibrium is the physically appropriate framework \cite{MITROVICA2004177}.

For a circular orbit, the orbital period $T$ can be derived from the centripetal acceleration equation and conservation of angular momentum:

\begin{equation}
\frac{v^2}{r} = \frac{GM}{r^2}
\end{equation}

\begin{equation}
L = mvr = \text{constant}
\end{equation}

These equations yield $v \propto \sqrt{G/r}$ and combined with $T = 2\pi r/v$, we obtain the scaling relation:

\begin{equation}
T_{\text{year}} \propto G^{-2}
\end{equation}

For a fluid body under constant angular momentum (fluid rotates nearly uniformly), fundamental scaling relations govern its response to changes in gravitational force \cite{Chandrasekhar:1987}. The radius $R$ of such a body scales with $G$ as:

\begin{equation}
R \propto G^{-1}
\end{equation}

This scaling arises from the balance between gravitational forces and pressure gradients in hydrostatic equilibrium. The moment of inertia $I$ then follows:

\begin{equation}
I \propto MR^2 \propto G^{-2}
\end{equation}

Conservation of angular momentum $L = I\omega$ (where $\omega$ is the angular velocity) then implies:

\begin{equation}
\omega \propto G^2
\end{equation}

Therefore, the length of a day (rotation period) would scale as:

\begin{equation}
T_{\text{day}} \propto \omega^{-1} \propto G^{-2}
\end{equation}

This crucial result shows that under fluid-body conditions, both $T_{\text{year}}$ and $T_{\text{day}}$ scale as $G^{-2}$. Consequently, the number of days per year $N = T_{\text{year}}/T_{\text{day}}$ would remain constant across any transition in $G$, consistent with the data shown in Figure~\ref{fig:dewinter} which is consistent with a constant number of days per year during the past 200Myrs (200 Mega-annum = 200Ma).

This conclusion differs fundamentally from analyses assuming Earth behaved as a rigid body with $R \propto G^0$, $I \propto G^0$, and $\omega \propto G^0$ \cite{Turcotte:2002}. Such scaling would lead to a discontinuity in the number of days per year across a $G$ transition. However, given Earth's more fluid-like state 150 Myr ago, supported by paleomagnetic data and mantle convection models \cite{Bercovici:2003}, the fluid body scaling relations are more appropriate for analyzing the $G$ transition period.

Under these conditions, the number of days in a year would remain unaffected by the transition, making the geochronological constraints discussed in Section 2.3 of \cite{Banik:2024yzi} not relevant for the GSM model. This conclusion is strongly supported by our current understanding of Earth's thermal and mechanical evolution \cite{Davies:2011} and is consistent with the continuous evolution of days per year shown in Figure~\ref{fig:dewinter}.

\section{Earth-Moon System}

The tidal interaction between Earth and Moon has played a fundamental role in the evolution of Earth's rotation rate \cite{Goldreich:1966, Murray:2000, Bills}. The tidal force between Earth and Moon follows an inverse cube law with distance:

\begin{equation}
F_{\text{tidal}} \propto \frac{GM_{\text{Moon}}M_{\text{Earth}}R_{\text{Earth}}}{R^3}
\end{equation}

where $R$ is the Earth-Moon distance and $R_{\text{Earth}}$ is Earth's radius. Since $R$ evolves to conserve angular momentum as Earth's rotation slows, we have $R \propto G^{-1}$. Therefore, the tidal stress scales as:

\begin{equation}
\sigma_{\text{tidal}} \propto \frac{F_{\text{tidal}}}{R_{\text{Earth}}^2} \propto G^5
\end{equation}
This is while considering Earth as a fluid body. On the extreme limit in the solid earth approximation, where $R_{Earth} \propto G^0$, the tidal stress scaling changes to $ \sigma_{\text{tidal}} \propto G^5$.
This strong dependence on $G$ in both scenarios means that a transition to lower $G$ would substantially weaken oceanic tides and reduce Earth's rotational deceleration rate \cite{Lambeck:1980, Banik:2024yzi}.

\begin{figure}[h]
\includegraphics[width=\columnwidth]{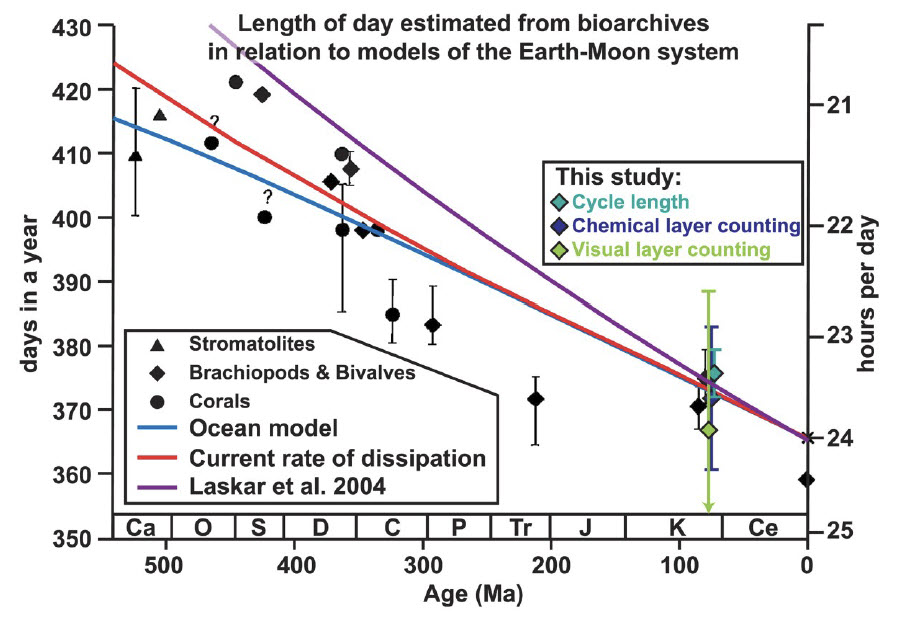}
\caption{Evolution of day length throughout the Phanerozoic based on various bioarchives (black symbols, from \cite{Williams:2000}) and new estimates from high-resolution chemical analyses of Torreites sanchezi rudist shells (green symbols). Three theoretical models are shown for comparison. Note the apparent reduction in the rate of day length increase during approximately the last 200 Myr, though measurement uncertainties (shown as error bars) remain significant. This pattern could potentially be interpreted as evidence for reduced tidal friction in recent geological time. Figure reproduced from \cite{deWinter:2020}.}
\label{fig:dewinter}
\end{figure}

The historical record of Earth's rotation rate shows interesting patterns, as illustrated in Figure \ref{fig:dewinter}. While measurement uncertainties are significant, the data suggests a reduced rate of change in day length during approximately the last 200 Myr. This is evident from the clustering of data points around 23-24 hours per day during this period, in contrast to the steeper increase in day length observed in earlier epochs.

However, interpreting this pattern requires careful consideration of multiple factors. The complex nature of Earth's rheological response to tidal forces, especially during periods when the mantle was less viscous, means that simple scaling relations may not fully capture the physics of tidal dissipation \cite{deWinter:2020, Webb:1982, Laskar:2004}. Recent studies have shown that mantle viscosity variations can significantly affect tidal dissipation rates \cite{Green:2017, Davies}.

Multiple sources contribute to uncertainties in these measurements, including challenges in preserving and interpreting daily growth bands in fossil records \cite{Wells:1963, Pannella:1972}, environmental effects on growth patterns \cite{Jones, Schone}, and technical limitations in measuring geological markers \cite{Berry:1968, Scrutton:1978}. Natural variability in Earth's rotation rate on various timescales further complicates interpretation \cite{Stephenson:2016, Morrison}.

Given these substantial uncertainties, the GSM model predicting a transition in $G$ at $z \approx 0.01$ cannot be ruled out based on tidal evolution arguments alone. In fact, the apparent reduction in Earth's rotational deceleration rate during the last 200 Myr could be viewed as tentatively supporting the GSM hypothesis, as a transition to lower $G$ would naturally lead to reduced tidal friction. However, this interpretation must be considered alongside other possible explanations and carefully weighed against the significant measurement uncertainties in the geological record.

While current uncertainties prevent a definitive conclusion, the apparent flattening in the rate of day-length increase over the past 200 Myr serves as a compelling hint that warrants further investigation. This test of the GSM could be made significantly more powerful with future geological studies. High-resolution cyclostratigraphy of sedimentary rock layers from the Cretaceous period (around 150-100 Myr ago) could map the evolution of Earth's rotational and orbital parameters with unprecedented precision. Such data could definitively confirm or rule out the smooth evolution of the day/year ratio during the proposed transition period, providing a sharp test of the fluid-Earth dynamics predicted by the GSM.

\section{Distance Indicators and Scaling Relations}
\subsection{Quantitative Analysis of Distance Measurements}

The comparison of distances measured using different stellar indicators provides a potential probe of modifications to gravity. Particularly important is the comparison between Cepheid variables and the Tip of the Red Giant Branch (TRGB) stars, as these are the primary calibrators of the cosmic distance ladder. As shown by \cite{Desmond:2020}, this comparison leads to surprisingly weak constraints on the strength of fifth forces or more general variations in the effective gravitational constant.

The key to deriving these constraints is that Cepheid and TRGB distances respond differently to modifications of gravity. For Cepheids, a strengthened gravitational force produces two effects:
\begin{enumerate}
    \item A reduction in the pulsation period scaling as $P \propto (G\rho)^{-1/2}$, leading to:
    \begin{equation}
    \Delta \log P = -\frac{1}{2}\log(1 + \Delta G/G_N)
    \end{equation}
    
    \item A potential increase in luminosity if the Cepheid's core is unscreened
\end{enumerate}

For TRGB stars, stronger gravity reduces their luminosity since the hydrogen-burning shell becomes unscreened. The TRGB distance modification follows:
\begin{equation}
\frac{D_{\rm true}}{D_{\rm GR}} = 1.021\left(1 - 0.04663\left(1 + \frac{\Delta G}{G_N}\right)^{8.389}\right)^{1/2}
\end{equation}

Since Cepheid-based distances are underestimated while TRGB distances are overestimated when standard GR-based analyses are applied, one might expect their comparison to provide strong constraints on gravity modifications. However, \cite{Desmond:2020} demonstrate through a detailed likelihood analysis that variations in the effective gravitational constant cannot be constrained to better than $\Delta G/G_N \sim 0.2$ at the 2$\sigma$ level using current data. This holds true whether the variation in $G$ comes from screened fifth forces or more general modifications to gravity.

Several factors contribute to these unexpectedly weak constraints:
\begin{itemize}
    \item Cepheids and TRGB stars may experience different effective gravitational constants as they occupy different regions within their host galaxies
    \item Combining distance measurements from different sources introduces significant systematic uncertainties
    \item Additional astrophysical effects may contribute to the scatter in distance estimates
\end{itemize}

The allowed magnitude of gravity modifications is thus much larger than might be expected from the $\sim$5\% precision typically quoted for individual distance measurements. This has important implications for theories involving screened fifth forces or other modifications to gravity, as it leaves significant room for such effects to impact the cosmic distance ladder. Future improvements in distance measurements with JWST and other facilities will be crucial for tightening these constraints by providing larger samples and better control of systematics.
\subsection{Systematic Uncertainties}
The comparison of different distance indicators as a probe of gravitational constant variations is subject to several important systematic uncertainties. These uncertainties can affect both the absolute calibration of individual distance indicators and their relative comparisons.

A fundamental systematic uncertainty arises from metallicity effects. The luminosity of stars depends not only on $G$ but also on their chemical composition. For Cepheid variables, metallicity affects both the zero-point and slope of the period-luminosity relation \cite{Romaniello}. While the TRGB magnitude is relatively insensitive to metallicity in the I-band \cite{Freedman:2020dne}, it shows stronger metallicity dependence in other passbands. The SBF method is particularly sensitive to stellar population properties, including age and metallicity variations \cite{Blakeslee_2012}.

Environmental effects present another source of systematic uncertainty. The local stellar environment can affect both the evolution and observability of distance indicators. Crowding and blending of stellar images become increasingly important at larger distances, potentially biasing luminosity measurements \cite{Yuan_2019}. Extinction and reddening corrections also introduce systematic uncertainties, particularly for Cepheid measurements in dusty star-forming regions \cite{Riess:2021jrx}.

The calibration of these distance indicators relies on primary anchors with geometric distances. The parallax measurements of Milky Way Cepheids by \textit{Gaia} have uncertainties of $\sim 1-2\%$ \cite{Riess:2021}, while the distance to the Large Magellanic Cloud, a crucial anchor for both Cepheid and TRGB calibration, is known to $\sim 1\%$ \cite{Pietrzynski:2019}. These calibration uncertainties propagate into all subsequent distance measurements.

When comparing different distance indicators to constrain variations in $G$, these systematic effects must be carefully considered. The total systematic uncertainty in the relative distance scales can be expressed as:

\begin{equation}
\sigma_{\rm sys}^2 = \sigma_{\rm cal}^2 + \sigma_{\rm met}^2 + \sigma_{\rm env}^2 + \sigma_{\rm ext}^2
\end{equation}

where $\sigma_{\rm cal}$, $\sigma_{\rm met}$, $\sigma_{\rm env}$, and $\sigma_{\rm ext}$ represent uncertainties from calibration, metallicity, environment, and extinction respectively. Current estimates suggest these systematic uncertainties limit the precision of $G$ constraints to $\sim 5\%$ \cite{Desmond:2019}, comparable to the statistical uncertainties from individual distance measurements.

Furthermore, potential correlations between different distance indicators could mask or mimic variations in $G$. For example, both Cepheids and TRGB stars in a given galaxy are subject to similar extinction and metallicity effects, potentially introducing correlated systematic errors in their distance estimates \cite{Freedman:2019}. This underscores the importance of using multiple independent distance indicators and cross-validation techniques to robustly constrain variations in fundamental constants.

\subsection{Evidence for Transitions in Distance Indicators}

Several analyses have identified potential transitions in distance indicator properties and fundamental constants at local cosmological scales. A comprehensive reanalysis of the SH0ES data by \cite{Perivolaropoulos:2022khd} has provided evidence for transitions at two characteristic distances: one at ~20 Mpc and another at ~50 Mpc. Their analysis introduced new degrees of freedom by allowing for transitions in four key parameters:
\begin{itemize}
    \item The SNIa absolute magnitude $M_B$
    \item The Cepheid absolute magnitude zero point $M_W$
    \item The Cepheid period-luminosity slope $b_W$
    \item The Cepheid metallicity-luminosity slope $Z_W$
\end{itemize}

For the transition at ~50 Mpc, the SNe Ia absolute magnitude shows a clear bifurcation:
\begin{equation}
M_B(D) = \begin{cases}
M_B^< = -19.25 \pm 0.03 \text{ mag} & \text{for } D < D_c \\
M_B^> = -19.43 \pm 0.15 \text{ mag} & \text{for } D > D_c
\end{cases}
\end{equation}
where $D_c \approx 50$ Mpc. This transition leads to a shift in the derived Hubble constant from $H_0 = 73.04 \pm 1.04$ km s$^{-1}$ Mpc$^{-1}$ to $H_0 = 67.33 \pm 4.65$ km s$^{-1}$ Mpc$^{-1}$, consistent with Planck/$\Lambda$CDM measurements. When including the inverse distance ladder constraint on $M_B^>$, the evidence strengthens with $\Delta\chi^2 \approx -15$ and the Hubble constant tightens to $H_0 = 68.20 \pm 0.88$ km s$^{-1}$ Mpc$^{-1}$.

Independent support for transitions comes from analyses of the Tully-Fisher relation. \cite{Alestas:2021} found hints of evidence at ~3$\sigma$ level hints for transitions at critical distances $D_c \approx 9$ Mpc and $D_c \approx 17$ Mpc. By splitting their sample at these distances, they identified tensions between subsamples at a level of $\Delta\chi^2 > 17$ (3.5$\sigma$). If interpreted as a gravitational transition, this would imply a decrease in the effective gravitational constant by $\Delta G/G \approx -0.1$ beyond these distances. While this change in $G$ has the anticipated magnitude, it occurs at somewhat lower distances than the transitions identified in the SH0ES analysis.

Several challenges exist in verifying these transitions:
\begin{itemize}
    \item Limited data in the critical transition regions (e.g., only 41 Cepheids and 4 SNe Ia beyond 50 Mpc in SH0ES)
    \item Peculiar velocities blurring the distance-time mapping
    \item Environmental effects that could mimic transitions
    \item Correlated systematic uncertainties between different distance indicators
    \item The 50 Mpc transition coinciding with the switch between calibrator and Hubble flow SNe Ia
\end{itemize}

The physical interpretation remains unclear, with several possibilities:
\begin{enumerate}
    \item Changes in fundamental constants affecting stellar physics
    \item Environmental effects on distance indicators
    \item Selection effects or observational biases
    \item Unrecognized calibration systematics
\end{enumerate}

Future observations will be crucial for testing these transition hypotheses:
\begin{itemize}
    \item More Cepheid-calibrated SNe Ia in the 40-60 Mpc range
    \item Alternative distance indicators like TRGB stars at transition distances
    \item Detailed studies of standardization parameters across transition regions
    \item Investigation of environmental dependencies
\end{itemize}

If confirmed, these transitions would have profound implications for both cosmology and stellar physics. Understanding their origin could provide insights into either fundamental physics or previously unknown systematic effects in distance measurements. The convergence of evidence from multiple independent analyses suggests this phenomenon warrants further investigation with future precision observations.

\section{Cosmic Chronometers and the GSM}

Cosmic chronometers measure $H(z)$  by comparing the age difference between older, passively evolving galaxies at slightly different redshifts \cite{Moresco:2018}. The basic equation relating the Hubble parameter to observable quantities is:

\begin{equation}
H(z) = -\frac{1}{1+z} \frac{\Delta z}{\Delta t}
\end{equation}

where:
\begin{itemize}
\item $H(z)$ is the Hubble parameter at redshift $z$, measured in units of \Hunit
\item $z$ is the redshift of the older galaxy in the pair being compared
\item $\Delta z$ is the redshift difference between the galaxy pair
\item $\Delta t$ is the difference in ages between the two galaxies at different redshifts
\end{itemize}

This can be derived from the fundamental relation in an expanding universe:
\begin{equation}
\frac{\dot{a}}{a} = H(z) = -\frac{\dot{z}}{1+z}
\end{equation}
where $a(t)$ is the cosmic scale factor normalized to $a(t_0)=1$ at the present time $t_0$, and dots denote derivatives with respect to cosmic time $t$. The negative sign appears because $z$ decreases as $t$ increases, following the relation $1+z = 1/a(t)$. For the small redshift differences used in cosmic chronometer measurements ($\Delta z \ll 1$), this differential relation can be approximated as:

\begin{equation}
H(z) \approx -\frac{1}{1+z} \frac{\Delta z}{\Delta t}
\end{equation}

This approximation becomes exact in the limit $\Delta z \to 0$, where the differential quotient becomes a derivative.

The measured age difference $\Delta t$ depends on stellar evolution timescales. For main sequence stars, the nuclear burning timescale $\tau$ scales with the gravitational constant $G$: as\cite{Adams:2008}:
\begin{equation}
\tau \propto L^{-1} \propto G^{-4}
\end{equation}

This  scaling has important implications for cosmic chronometer measurements. The strong inverse dependence on $G$ means that even small variations in the gravitational constant can lead to significant changes in stellar evolution rates.

This leads to the relation between measured and true quantities:
\begin{equation}
\Delta t_{\mathrm{measured}} = \Delta t_{\mathrm{true}} \left(\frac{G}{G_0}\right)^{-4}
\end{equation}

Therefore, the measured Hubble parameter relates to the true value as:
\begin{equation}
H_{\mathrm{measured}}(z) = H_{\mathrm{true}}(z) \left(\frac{G}{G_0}\right)^4
\end{equation}

For small variations in $G$, we can linearize this relation:
\begin{equation}
H_{\mathrm{measured}}(z) \simeq H_{\mathrm{true}}(z) \left(1 + 4\frac{\Delta G}{G_0}\right)
\end{equation}

Thus, the true Hubble parameter can be recovered from the measured value via:
\begin{equation}
H_{\mathrm{true}}(z) = \frac{H_{\mathrm{measured}}(z)}{1 + 4\Delta G/G_0}
\end{equation}

For the specific case where $\Delta G/G_0 = 0.04$ (i.e., $G$ was 4\% larger in the past), we have:
\begin{equation}
H_{\mathrm{true}}(z) = \frac{H_{\mathrm{measured}}(z)}{1 + 4(0.04)} = \frac{H_{\mathrm{measured}}(z)}{1.16}
\end{equation}

This 16\% correction is particularly interesting because it could lead to a lower true value of $H_0$ even if a high value consistent with SH0ES distance ladder result was measured by CC. Thus, this discrepancy could be explained if CC measurements appear artificially high due to larger $G$ in the past, while the true expansion history follows the Planck $\Lambda$CDM prediction.

This is particularly relevant as most current cosmic chronometer measurements of $H_0$ tend to fall in an intermediate range between the Planck and SH0ES values, with some analyses favoring higher values consistent with the distance ladder. The 16\% correction derived above is significant enough to reconcile a CC-inferred value of $H_0 \approx 73$ \Hunit~with the Planck value of $H_0 \approx 67.4$ \Hunit~($73/1.16 \approx 63$). Therefore, the GSM provides a clear physical mechanism for reinterpreting the cosmic chronometer data, suggesting that their measurements may be systematically high due to a failure to account for a variation in the gravitational constant.

\section{Discussion and Conclusions}

\subsection{Summary of Quantitative Uncertainties}

Our detailed analysis has revealed that many of the apparent challenges to the GSM are significantly weakened when proper uncertainties are considered. The CMB and BAO constraints from recent analyses allow for variations in the gravitational constant within the range $0.976 \lesssim G/G_0 \lesssim 1.065$ at the 2$\sigma$ level \cite{Lamine:2024xno}. This range becomes even more permissive when considering systematic uncertainties in recombination physics and screening mechanisms \cite{Uzan:2011}.

In the domain of stellar evolution, modern calculations have substantially revised our understanding of the luminosity-gravity relationship. The work of \cite{Adams:2008} demonstrates that stellar luminosity scales approximately as $L_\odot \propto G^4$, rather than the traditional Teller scaling of $G^7$. This modification significantly reduces the tension between the GSM and stellar evolution constraints. Under this revised scaling, a 5\% increase in $G$ would result in a 20\% increase in stellar luminosity, leading to an approximately 0.8 Gyr shift in apparent stellar ages. These modifications are well within the uncertainties of current stellar modeling techniques \cite{Choi:2016}.

The relationship between Type Ia supernova luminosity and gravitational constant has been extensively studied, with work by \cite{Wright:2017} establishing important constraints on possible variations. The comprehensive analysis of \cite{Scolnic:2018} provides crucial systematic uncertainty estimates through empirical cross-validation studies.

Distance indicator comparisons provide another crucial test of the GSM. Analyses by \cite{Desmond:2020} demonstrate that current observations allow for variations in $G$ at the level of $\Delta G/G_N \sim 0.2$ at 2$\sigma$ confidence, a much wider range than previously assumed. This flexibility in the distance scale provides important breathing room for gravitational modifications of the type proposed in the GSM.

\subsection{Future Research Directions}

Observational facilities will play a crucial role in testing the GSM through multiple complementary approaches. The combination of precision distance measurements \cite{Riess:2021} with detailed studies of stellar evolution under modified gravity scenarios \cite{Sakstein:2015} provides a promising path forward.

Theoretical work must advance in parallel with these observational capabilities. The development of more sophisticated stellar evolution models incorporating screening mechanisms and modified gravity effects represents a crucial frontier \cite{Burrage:2018}. These models must account for the complex interplay between gravitational physics and other fundamental forces in stellar interiors, as highlighted by \cite{Capozziello:2011}.

Environmental effects present another important avenue for future research. The work of \cite{Desmond:2019} suggests that spatial variations in effective gravitational strength might be correlated with large-scale structure. Understanding these correlations could provide new tests of the GSM and help distinguish it from other proposed solutions to the Hubble tension.

\subsubsection*{A Roadmap for Future Tests}

The G-step Model (GSM) makes specific, falsifiable predictions that can be rigorously tested with next-generation observational facilities. We propose the following roadmap to either confirm or rule out the model:

\begin{itemize}
    \item \textbf{Mapping the Transition with JWST:} A targeted survey with the James Webb Space Telescope (JWST) to discover and obtain high-precision photometry for more than 20 new Cepheid-calibrated Type Ia supernovae in the crucial 40-60 Mpc distance range. The discovery of a systematic shift in the absolute magnitude ($M_B$) of approximately -0.18 mag beyond $\sim$50 Mpc would provide definitive evidence for the transition predicted by the GSM.
    \item \textbf{Cross-Validation with TRGB:} A parallel effort to calibrate the distances to supernova host galaxies in the same 40-60 Mpc range using the Tip of the Red Giant Branch (TRGB) method. As the G-dependence of TRGB luminosity differs from that of Cepheids and SNe Ia, this would provide a crucial cross-check to distinguish a physical transition in $G_{\text{eff}}$ from unknown astrophysical systematics.
    \item \textbf{Probing Gravity with Standard Sirens:} Future gravitational wave observatories, such as the Einstein Telescope, will be able to measure the luminosity distance to binary neutron star mergers with high precision. Comparing these "standard siren" distances to the electromagnetic distances of their host galaxies in the transition region can directly probe for a change in the effective gravitational constant, as the propagation of gravitational waves is sensitive to $G_{\text{eff}}$.
    \item \textbf{Constraining Stellar Evolution:} High-precision asteroseismology of solar-type stars in open clusters with a range of ages could provide tighter constraints on the luminosity-gravity scaling relation ($L \propto G^n$). Confirming that the power-law index $n$ is closer to 4 than to 7 would further solidify a key argument in favor of the GSM's viability.
\end{itemize}

\subsection{Model Viability Assessment}

Our comprehensive review indicates that the GSM remains a viable solution to the Hubble tension when proper uncertainties are considered. The model's consistency with CMB constraints at the 2$\sigma$ level, particularly when accounting for systematic uncertainties in early universe physics, represents a significant success \cite{Galli:2009, Lamine:2024xno}. The revised understanding of stellar evolution under modified gravity, primarily through the work of \cite{Adams:2008}, has substantially reduced the tension with stellar physics constraints.

The evidence for transitions in distance indicators at characteristic scales provides intriguing support for the GSM framework. Independent analyses by \cite{Perivolaropoulos:2021} and \cite{Alestas:2021} have identified potential transitions in various distance indicators at scales consistent with GSM predictions. These results align with detailed investigations of galaxy scaling relations \cite{McGaugh:2020} and the cosmic distance ladder \cite{Freedman:2021}.

The GSM offers several theoretical advantages over alternative solutions to the Hubble tension. By proposing modifications only to local physics, it preserves the remarkable success of $\Lambda$CDM in describing the early universe and large-scale structure formation. The model naturally explains the apparent discrepancy between distance ladder and other $H_0$ measurements, as demonstrated by \cite{Mortsell:2021}. Furthermore, it provides a physical mechanism for observed transitions in various distance indicators, connecting seemingly disparate astronomical phenomena within a unified framework \cite{Verde:2019}.

A potential criticism of the G-step Model (GSM) is that a sharp transition at the specific redshift of $z \approx 0.01$ appears fine-tuned. We argue, however, that this feature should be viewed in the context of the problem it aims to solve. The Hubble tension is itself a highly ``fine-tuned'' issue, representing a precise, high-significance discrepancy between two otherwise robust measurement classes. It is therefore plausible that its resolution may not come from a generic, slowly-evolving mechanism, but rather from a new physical phenomenon that becomes active at a specific, recent cosmological epoch. Thus, we interpret the proposed transition not as an arbitrary fix, but as a potential signature of new physics, such as a late-time phase transition, manifesting at a particular energy scale.

While the GSM is phenomenological, viable physical mechanisms capable of producing such an ultra-late-time transition have been proposed. A leading candidate is a first-order phase transition, where our local region of the Universe has recently transitioned from a metastable (false) vacuum to a new, true vacuum state. A concrete realization of this scenario was explored by Efstratiou and Perivolaropoulos \cite{Efstratiou:2025xou}, who investigated the decay of a metastable vacuum in the context of scalar-tensor theories. In such models, the transition occurs via the nucleation of ``gravitational bubbles''---regions of space with a different true vacuum value for the scalar field, which in turn leads to a different value for the effective gravitational constant $G_{\text{eff}}$. Notably, their analysis shows that a transition occurring around $z\approx0.01$ would produce bubbles with an observable radius of $\sim$50 Mpc today, providing a well-motivated physical basis for the transition scale required to resolve the Hubble tension.

While challenges remain, particularly in developing a complete theoretical framework for the transition mechanism, none of the current observational constraints definitively rule out the GSM. The possibility that we live in a local region with modified gravitational physics represents a testable hypothesis that merits continued investigation.

\section*{Acknowledgements}
This research was supported by COST Action CA21136 - Addressing observational tensions in cosmology with systematics and fundamental physics (CosmoVerse), supported by COST (European Cooperation in Science and Technology). The Author R is supported by Project SA097P24, funded by Junta de Castilla y Leon. R would also like to thank Southampton University for the hospitality and resources, where part of this project was done.
\bibliographystyle{apsrev4-2} 
\bibliography{references}

\end{document}